\documentclass[a4paper,english,12pt]{article}

\usepackage{amssymb}
\usepackage{amsmath}
\usepackage{overcite}
\makeatletter \renewcommand\@biblabel[1]{$^{#1}$} \makeatother

\newcommand{\proof}{\noindent{\it Proof.}\ \ }
\newcommand{\qed}{\hfill $\Box$}
\newcommand{\R}{\mathbb{R}}
\newcommand{\RR}{\mathbb{R}^+_0 \! \times \mathbb{R}}
\newcommand{\ol}[1]{\overline{#1}}
\newcommand{\D}{\mbox{div}}

\newtheorem{thm}{Theorem}
\newtheorem{prop}{Proposition}
\newtheorem{rem}{\it Remark}

\begin{document}

\title{Relativistic stars in differential rotation: bounds on the
{\em dragging rate} and on the {\em rotational energy}}
\author{M. J. Pareja\\
Dept. de F\'{\i}sica Te\'orica II, Facultad de Ciencias F\'{\i}sicas,\\
Universidad Complutense de Madrid, E--28040 Madrid, Spain\\
{\em e-mail: mjpareja@fis.ucm.es}
}
\date{}

\maketitle
\begin{abstract}
\noindent For general relativistic equilibrium stellar models (stationary axisymmetric asymptotically flat
and convection-free) with differential rotation, it is shown that for a wide class of rotation laws
the distribution of angular velocity of the fluid has a sign, say ``positive", and then both the dragging rate
and the angular momentum density are positive.
In addition, the ``mean value" (with respect to an intrinsic density) 
of the dragging rate is shown to be less than the mean value of the fluid angular velocity
(in full general, without having to restrict the rotation law, 
nor the uniformity in sign of the fluid angular velocity); this inequality yields
the positivity and an upper bound of the total rotational energy.
\end{abstract}


\section*{\normalsize I. INTRODUCTION}

One of the most interesting of the relativistic effects produced by the rotation of a star
is the dragging of inertial frames (also called Lense-Thirring effect)\cite{Thirring}.
This has been classically described in terms of its {\em local} effects on gyroscopes and
particles. However, its {\em cumulative} effects on the motion of particles give a much
simpler description: a locally non-rotating test particle, that is dragged along in the 
gravitational field of the star, has an angular velocity, as seen from a non-rotating 
observer at spatial infinity, which is named {\em angular velocity of cumulative dragging}, 
or {\em rate of rotational dragging}, shortly called {\em dragging rate} \cite{Thorne,Bardeen}. 
It is physically expected that, for isolated rotating stars in thermodynamic equilibrium, 
this dragging rate $A$ has the same sign (rotation sense) as the fluid angular velocity $\Omega,$
if this one has a ``uniform'' sign throughout the fluid (in the general differentially rotating case). 
Indeed, Lindblom\cite{Lindblom} and, independently, Hansen and Winicour~(1977) \cite{H-W} seem
to establish this result, however, without explicitly fulfilling the corresponding requirements
when applying the called Hopf theorem (a maximum principle) to an elliptic operator in a certain domain, 
concerning the boundedness of its coefficients on the boundary of the domain, specifically on the
axis of rotation, and the $C^1$ (and not $C^2)$ regularity of the metric functions across the surface
of the star.

Also, (assuming in the description above that the test particle does not collide with the star's 
matter if it goes through the star) one is tempted to conjecture, in principle in the rigidly 
rotating case with $\Omega =$ const.$\, = \Omega_{\ast} > 0,$ that the dragging rate is
bounded above by the fluid angular velocity, $A \le \Omega_{\ast}.$  
And Hansen and Winicour~(1975) \cite{H-W} offer some proof of this (although with the same objection
as above).

In the general differentially rotating case, however, an analogous relation 
should not be expected to be so simple; firstly, because different portions of the star's interior
could have opposite rotational motion about the same axis (assuming a convection-free fluid),
and, secondly (even if the fluid angular velocity has a sign), because, due to the integrability 
condition of the equation of motion, the distribution of fluid angular velocity,
$\Omega$-profile, cannot be freely prescribed; instead, it is derived (together with the potential
functions, integrating the field equations) once an appropriate {\em rotation law} is given. 
Most of the literature concerning
numerical works on differentially rotating neutron stars make generally the ansatz for
a certain rotation law which yields $\, A \le \Omega.$ Nevertheless, 
for a more general law such a relation is not so obvious. 
Hansen and Winicour~(1977) \cite{H-W} have made some attempts to give a result, 
however they needed the unphysical assumption that the star's matter occupies the whole space.

One of the aims of this work is precisely to find general and physically reasonable assumptions
on the rotation law of a differentially rotating stellar model, so that the dragging rate
is (at each interior point) less than the fluid angular velocity, and, hence, the {\em angular
momentum density} is positive (vanishing on the axis) provided the weak energy condition 
is satisfied. For that matter we consider the time-angle field equation's component, which is
elliptic and linear in the dragging potential $A$ in coordinates adapted to the symmetries.
The approach with the metric in these coordinates 
is attractive because the field equations become semilinear elliptic.
Specially, they reduce to four (coupled) elliptic equations for the
four metric functions. One has however to control
the coordinate singularities of the equations on the axis of rotation,
but these can be treated mathematically using the axial symmetry of
the physical problem \cite{Sch}. So handled, the elliptic equation in $A$ writes in a ``regular''
form and has bounded coefficients; this allows us to apply a maximum principle to several
differential inequalities, which, using the $C^1$-matching on the star's surface and the 
asymptotic flatness of the metric, will lead to the mentioned and other interesting
inequalities.

\bigskip

More generally, as was conjectured by Thorne\cite{Thorne} (p.~245), 
the {\em mean value} of $A$ is expected to be less than the 
{\em mean value} of $\Omega.$ However, to my knowledge, there is in the literature no 
explicit and so general result in this direction. The other purpose of the present work is then 
to derive a ``general'' inequality on ``mean values'' with respect to a density function. 
In addition, related to this question is the concept of {\em total rotational energy} of the star.
Hartle\cite{H4} has given bounds for this rotational energy in the slow rotation limit, which
we aim here to generalize. 

\bigskip

The paper is organized as follows. After reviewing in Sec.~II the model for a relativistic star
which is rotating differentially, Sec.~III is devoted to handle 
the concerned field equation, elliptic and linear in the dragging potential, with special
attention to the regularity and boundedness properties of the involved functions, as a
preparation allowing us to apply the maximum principles (reviewed in the Appendixes)
in Sec.~IV, where inequalities concerning mainly the positivity of the dragging rate and
of the angular momentum density are derived. In Sec.~V a general ``mean values inequality'' 
is derived in full general; and the positivity and upper bound of the total rotational energy
is established in the general differentially rotating case. Finally, in Sec.~VI the relevant
results are briefly summarized.

\section*{\normalsize II. MODEL FOR A DIFFERENTIALLY ROTATING RELATIVISTIC STAR}

The spacetime of an isolated rotating star in thermodynamic equilibrium within
general relativity theory is generally represented by an asymptotically flat
stationary axisymmetric $4$-dimensional Lorentzian manifold $(\cal{M},{\mathbf g}),$
with metric ${\mathbf g} = g_{\alpha \beta} dx^{\alpha} dx^{\beta}$ satisfying
Einstein's equations,
\begin{equation}
\label{eq:Einstein}
G_{\alpha \beta} := R_{\alpha \beta} - \frac{1}{2} R \, g_{\alpha \beta} = 
8 \pi \, T_{\alpha \beta}
\, ,
\end{equation}
for the energy-momentum tensor of a perfect fluid, 
$T_{\alpha \beta} = (\varepsilon + p) \, u_{\alpha} u_{\beta} + p \, g_{\alpha \beta},$
with $4$-velocity $u^\alpha,$ energy density $\varepsilon,$ and pressure $p.$ 
Signature of ${\mathbf g}$ is here considered to be $(- + + +).$
Since the star is isolated, 
the matter (perfect fluid) is confined in a compact region in the space
(interior), with vacuum, $T_{\alpha \beta} = 0,$ on the outside.

We denote the two (commuting) global time and axial Killing vector fields \cite{C} by
${\boldsymbol \xi} = \partial_t$ and ${\boldsymbol \eta} = \partial_{\phi},$ 
respectively, where $x^0 \equiv t$ labels the space-like hypersurfaces which are
invariant under time translations, and $x^1 \equiv \phi$ is the axial-angle coordinate
around the axis of rotation, given by ${\boldsymbol \eta} \equiv 0;$ 
\, $(t,\phi) \in \R \! \times \! [0,2 \pi[.$ The metric
components will then only depend on the two remaining spatial coordinates,
$g_{\alpha \beta} = g_{\alpha \beta}(x^2, x^3).$

We shall assume that the fluid motion is purely azimuthal (non-convective), 
i.e.~the fluid $4$-velocity is contained in the $2$-surface spanned by the two
Killing fields, (as $1$-forms) 
\begin{equation}
\label{eq:circul}
{\mathbf u} \wedge {\boldsymbol \xi} \wedge {\boldsymbol \eta} = 0 \hspace{2cm}
\mbox{(circularity condition)}
\, .
\end{equation}
In that case it can be seen \cite{K-T} that the $2$-surface elements orthogonal
to the $2$-dimensional group orbits of the Killing fields are surface forming
(the same holds in the vacuum region); and, consequently, the metric may be written 
in a form which is explicitly symmetric under the change 
$(t,\phi) \rightarrow (-t,-\phi).$ In the $2$-surfaces orthogonal to the orbits
we can always introduce {\em isotropic coordinates} $(x^2, x^3) = (\rho,z)$
without loss of generality, so that the metric can always be reduced to the
standard form \cite{Kramer,Bardeen} 
\begin{equation} 
\label{metric}
{\mathbf g} =
g_{\alpha \beta}dx^{\alpha} dx^{\beta} = 
    -e^{2U}dt^{2}+
e^{-2U}\left[\rho^2 e^{2B} (d\phi-A\,dt)^{2}+
    e^{2K}(d\rho^{2} +
dz^{2})\right] 
\, ,
\end{equation}
where the metric functions $K, U, B,$ and $A$ depend only on the $(\rho,z)$-coordinates 
of the ``meridian plane''. Here $\rho$ and $z$ are cylindrical coordinates
at the asymptotically flat infinity, and, using the remaining freedom of conformal
transformations in the meridian plane, we choose these coordinates such that
$\rho = 0$ represents the axis of rotation and $(\rho,z) \in \RR$
(denoting $\R^+_0 := \{ \mbox{x} \in\R\,|\, \mbox {x} \ge 0\}).$
The metric functions $\rho \, e^B, \, U,$ and $A$ can be written as invariant combinations of 
the Killing fields in the form
\\
\parbox{13cm}{
\begin{eqnarray*}
\rho^2 e^{2B} &=&
-\det((g_{\mu \nu})_{\mu,\nu =t,\phi}) =
            -{\mathbf g}({\boldsymbol \xi},{\boldsymbol \xi})\,
{\mathbf g}({\boldsymbol \eta},{\boldsymbol \eta})
+{\mathbf g}({\boldsymbol \xi},{\boldsymbol \eta})^{2} \\ 
e^{2U} &=&  \frac{\rho^2 e^{2B}}{{\mathbf g}({\boldsymbol \eta},{\boldsymbol \eta})}
\\
A      &=&  -\frac{{\mathbf g}({\boldsymbol \xi},{\boldsymbol \eta})}
{{\mathbf g}({\boldsymbol \eta},{\boldsymbol \eta})}
\; ,
\end{eqnarray*}
}
\hfill
\parbox{1cm}{\begin{equation}\label{def:BUA}\end{equation}}
\\
and they can be interpreted physically as follows:
$\rho \, e^B$ represents a sort of distance from the rotation axis (and, hence, 
$B$ is, to some extent, a measure how far is $\rho$ from being that distance); $U$ is
a generalization of the gravitational potential; and $A$ is the 
{\em angular velocity of cumulative dragging}, or {\em dragging rate}.
The remaining metric function is $K,$ the conformal factor in the meridian plane.

\vspace{2mm}

Throughout the following we shall denote the closure and the boundary 
of a set $X$ by $\overline{X}$ and $\partial X,$ respectively.
We fix the notions 
\begin{equation}
\label{def:EIS}
\parbox{11.5cm}{
\parbox{10cm}{
\begin{eqnarray*}
I & \equiv  & \mbox{interior of the star} := 
\{(\rho,z) \in \RR \ | \ p(\rho,z)>0\}
         \subset \RR \\
E & \equiv  & \mbox{exterior of the star} := 
(\RR)\setminus\ol{I}\subset \RR \\
S & \equiv  &
\mbox{star's surface} := \ol{I}\cap\ol{E} 
        = \partial I 
        \subset \RR \, ,
\end{eqnarray*}
}}
\end{equation}
$I$ and $E$ open in the {\em induced topology in} $\RR \subset \R^2;$ 
that means, although part of
the axis $(\rho = 0)$ is in $I$ (and part in $E),$ the only points of the axis
which are in $\partial I = S$ (and in $\partial E = S \cup \{\infty\})$ 
are the poles, if they exist. The set $I \subset \RR$ is supposed to be
bounded and connected. Concerning the regularity of $S = \partial I,$ we assume
it satisfies an exterior sphere condition everywhere (cf.~Definition in Appendix~A).

\vspace{2mm}

Within our star model, the {\em matching} conditions 
(from the interior and the exterior solutions) require that the pressure 
vanishes identically on the star's surface, $p=0$ on $S.$ In the exterior
$(T_{\alpha \beta} = 0)$ we have $\varepsilon = p = 0.$
Furthermore, $\varepsilon$ and $p$ satisfy a barotropic 
equation of state in the interior, 
\begin{equation}
\label{eq:eos}
\varepsilon = \varepsilon(p) \quad \mbox{in} \ \ol{I}
\, .
\end{equation}
We assume the pressure $p$ to be continuous with respect to the coordinates, 
and also $p \mapsto \varepsilon(p)$ a continuous function, 
\begin{equation}
\label{cond:cont_p}
p \in C^0(\RR) \, , \qquad
p \mapsto \varepsilon(p) \, \in C^0(\R^+_0)
\, ,
\end{equation}
satisfying the {\em weak energy condition} \cite{Wald}, 
\begin{equation}
\label{eq:energy-cond}
\varepsilon + p \ge 0  \qquad (\mbox{in} \ \, \RR) \, .
\end{equation}
(Notice, by the definition of the interior, (\ref{def:EIS}), 
if $\varepsilon \ge 0$ in $\ol{I},$ as it is generally assumed, 
we shall have even $\varepsilon + p > 0$ in $I,$ and, hence, 
condition (\ref{eq:energy-cond}) follows. In addition, since the equation
of state is defined only in the interior, (\ref{eq:eos}), requirement 
(\ref{cond:cont_p}) does not guarantee the continuity of $\varepsilon$
across the star's surface (where $p=0),$ namely, if 
$\varepsilon (p=0) > 0,$ then a jump discontinuity of $\varepsilon$ across the star's 
surface occurs.)

\vspace{2mm}

From the circularity condition (\ref{eq:circul}) on the fluid $4$-velocity 
(in $\ol{I}),$ this is of the form 
$$
{\mathbf u} = u^t ({\boldsymbol \xi} + \Omega {\boldsymbol \eta}), 
\ \ \ \ \mbox{where} \ \ \Omega \equiv \frac{u^{\phi}}{u^t} =
\frac{d\phi}{dt}
$$
is the angular velocity of the fluid measured by a distant observer in an 
asymptotically flat spacetime, and
the fact that the $4$-velocity ${\mathbf u}$ is a unit time-like vector field
determines the normalization factor $u^t,$ such that \ 
${\mathbf g}({\mathbf u},{\mathbf u}) = -1,$ 
i.e.
\begin{equation}
\label{def:N}
(u^t)^{-2} = e^{2U} - \rho^2 e^{2(B-U)}(\Omega - A)^2 \, =: \, N 
\, ,
\end{equation}
from where $N = (u^t)^{-2} > 0$ in $\ol{I}.$ Indeed, we do not allow 
that the velocity of light is approached somewhere, and, hence, even 
\begin{equation}
\label{N>0}
N \ge \mbox{const.} > 0 \quad \mbox{in} \ \ol{I}.
\end{equation}
We consider a star rotating differentially with a distribution of angular velocity 
({\em rotation profile}) $\Omega = \Omega(\rho,z),$ a continuously differentiable function, 
\begin{equation}
\label{cond:Om_C1} 
\Omega \in C^1 \! \left(\ol{I} \right) 
\, .
\end{equation}
However, the $\Omega$-profile of the fluid cannot be freely chosen, this shows up in 
the following. The integrability conditions of the field equations (\ref{eq:Einstein}),
that is, the equation of hydrostatic equilibrium $T^{\alpha \beta}_{\ \ \ ;\beta} = 0$
(from $G^{\alpha \beta}_{\ \ \ ;\beta} = 0)$ (where `${}_{;}$' denotes covariant derivative),
particularly, its part orthogonal to the fluid $4$-velocity ${\mathbf u},$ 
reduces to the Euler equation, 
\begin{equation}
dp = - (\varepsilon + p)\, {\mathbf a} \, ,
\label{eq:Euler}
\end{equation}
where $\mathbf{a}$ is the $4$-acceleration of the fluid,
$\mathbf{a} = \nabla_{{\mathbf u}} {\mathbf u}.$ Specifically, 
\begin{equation}
{\mathbf a} = dV + u^t u_{\phi} \, d \Omega \, , \qquad V \equiv \frac{1}{2} \ln N
\, ,
\label{def:acc}
\end{equation}
$u^t u_{\phi} = \rho^2 e^{2(B-U)} (\Omega - A) N^{-1}.$ But the integrability condition
of Eq.~(\ref{eq:Euler}) taking into account (\ref{eq:eos}) is \ $d {\mathbf a} = 0;$
following then, from (\ref{def:acc}), $d (u^t u_{\phi}) \wedge d \Omega = 0.$ 
The special case where $\Omega=$ const. is called {\em rigid rotation} 
(or {\em uniform rotation}). In general we shall have $\Omega \neq$ const.,
following then, 
\begin{equation}
\label{cond:dr}
u^t u_{\phi} = {\cal F}(\Omega) 
\, ,
\end{equation}
for some function ${\cal F},$ {\em rotation law}. By specifying the function ${\cal F}(\Omega)$ 
a specific model of {\em differential rotation} is obtained. 
(Note, since in the Newtonian limit $\, u^t u_{\phi} \to \rho^2 \, \Omega,$ 
Eq.~(\ref{cond:dr}) expresses the general relativistic generalization of the Newtonian 
``rotation on cylinders'' theorem, $\Omega = {\cal G}(\rho^2) \, $).

\vspace{4mm}

Further requirements on our stellar model are:
\begin{enumerate}

\item[a.] the metric functions are (at least) two times continuously differentiable 
in the interior and in the exterior of the star, and continuously differentiable 
everywhere (cf.~Note in Subsec.~III.B),
\begin{equation}
\label{cond:C1C2}
K, U, B, A \, \in \, C^2(I) \cap C^2(E) \cap C^1(\RR) \, ;
\end{equation}

\item[b.] in order that the metric functions are symmetric with respect to the 
$z$-axis $(\rho = 0)$ (``axisymmetric solutions''), and, hence, the metric (\ref{metric}), 
defined on $\cal{M}$ excluding the axis, can be extended to an at least $C^1$ 
axisymmetric tensor field in the whole spacetime $\cal{M},$ we assume that
\begin{equation}
\label{cond:axisym}
\mbox{as} \ \rho \to 0, \qquad \partial_{\rho} K, \ \partial_{\rho} U, \ 
\partial_{\rho} B, \ \partial_{\rho} A \to 0 \, ,
\end{equation}
and, for completeness, also \ $\partial_{\rho} \varepsilon, \, \partial_{\rho} p \to 0 \, ;$

\item[c.] finally, by the asymptotic flatness requirement,
denoting ${\cal D} := (\partial_{\rho}, \partial_z),$
\begin{equation}
\label{cond:as-flat}
\mbox{as} \ \mbox{\it \small R} := ({\rho}^2 + z^2)^{1/2}\to\infty, \quad 
K, U, B, A \to 0 \quad \mbox{and} \quad
{\cal D} K, \, {\cal D} U, \, {\cal D} B, 
\, {\cal D} A \to {\mathbf 0} 
\, .
\end{equation}

\end{enumerate}

Notice, from $C^1$ regularity, in~(\ref{cond:C1C2}), and asymptotic flatness,
(\ref{cond:as-flat}),
it follows, in particular, that
the metric functions and their derivatives
are bounded \cite{fn1},
\begin{equation}
\label{cond:bound0}
|K|, |U|, |B|, |A| < \infty \quad \mbox{and} \quad 
\|{\cal D} K\|, \, \|{\cal D} U\|, \, \|{\cal D} B\|, \, \|{\cal D} A\| \, 
< \infty \ \ \mbox{in} \ \RR \, .
\end{equation}

\section*{\normalsize III. THE ELLIPTIC EQUATION FOR THE DRAGGING RATE $A$}

\subsection*{A. The time-angle field equation component}

The $(t \phi)$ component of Einstein's equation (\ref{eq:Einstein}) 
in these coordinates takes the form \cite{Sch}
\begin{eqnarray}
& & \partial_{\rho \rho} A + \partial_{zz} A + \frac{3}{\rho} \, \partial_{\rho} A +
\langle 3 {\cal D}B - 4 {\cal D}U \, , \, {\cal D}A \rangle = - \psi^2 \cdot (\Omega - A) \, ,
\label{eq:A-eq0} \\
\mbox{with} & & \psi^2 := 16 \pi \frac{e^{2K}}{N} (\varepsilon + p) \ \ \
(\ge 0, \ \ \mbox{by} \ (\ref{eq:energy-cond}) \ \mbox{and} \ (\ref{N>0})) 
\label{psi2}
\, ,
\end{eqnarray}
where $\langle . , . \rangle$ denotes the Euclidean scalar product. 
Since, from condition (\ref{cond:axisym}), $v(\rho,z) = v(-\rho,z)$ for $v=K,U,B,A,$ \ 
i.e.~we are considering only axisymmetric solutions of the field equations,
and since only ``axisymmetric operations''
appear in these equations, we consider the following transformation
(in the spirit of Ref.~6) in order to avoid the coordinate singularity
(of Eq.~(\ref{eq:A-eq0})) on the axis of symmetry $(z$-axis, i.e.~$\rho = 0).$
To this end we use the $5$-{\em lift of} each function $v \equiv v(\rho,z)$ 
(on $\R^5),$ for the metric functions $v = K,U,B,A$ and also for
$v = \Omega, \varepsilon, p,$ where the $n$-{\em lift of}
$v: \! \RR \rightarrow \R$ on flat $\R^n,$ axisymmetric around the
$x_n$-axis, is defined as follows
\begin{equation}
\label{def:n-lift}
v \mapsto \tilde{v} \quad \mbox{such that} \quad 
\tilde{v}(x) \equiv \tilde{v}(x_1,\ldots,x_n):=
v \! \left( \rho = (x_1^{\ 2} + \cdots + x_{n-1}^{\ 2})^{1/2} \, , 
\, z = x_n \right) \, ,
\end{equation}
and, for every function $\tilde{v}: \R^n \rightarrow \R,$ 
the {\em meridional cut (in direction $x_1$) of} $\tilde{v},$
\begin{equation}
\label{def:merid_cut}
\tilde{v} \mapsto v \qquad \mbox{such that} \qquad 
v(\rho,z):= \tilde{v}(\rho,0,\ldots,0,z) \, .
\end{equation}
For axisymmetric functions, both operations are isomorphisms and inverse
to each other; but the relevant properties of $n$-lift and meridional cut
are that $(a)$ they leave the regularity conditions and the norms invariant,
$(b)$ they commute with ``axisymmetric operations'', in particular, with all
operations in Eq.~(\ref{eq:A-eq0}), like multiplication and scalar product,
yielding specially (for $n=5)$
$$
\langle {\cal D}v , {\cal D}w \rangle = \langle \nabla \tilde{v} , \nabla \tilde{w} \rangle
 \, ,
\quad \mbox{denoting} \ \nabla := (\partial_1, \ldots, \partial_5) 
\qquad (\partial_i \equiv \partial_{x_i}, 
\ \ \partial_{ij} \equiv \partial_{x_i} \partial_{x_j} \ )
\, , 
$$
and, remarkably, $(c)$ they transform the operator 
$\partial_{\rho \rho}  + \partial_{zz}  + \frac{n-2}{\rho} \partial_{\rho}$ \, 
$(n \ge 2)$ into the flat $n$-dimensional Laplacian, and vice versa; having 
for $n=5$ 
$$
\partial_{\rho \rho} v  + \partial_{zz} v  + \frac{3}{\rho} \partial_{\rho} v = 
\sum_{i=1}^5 \partial_{ii} \tilde{v} =: \Delta \tilde{v}
\, . 
$$
Hence, with the $5$-lift, Eq.~(\ref{eq:A-eq0}) writes in the form
\begin{equation}
\label{eq:A-eq}
\Delta \tilde{A}+
\langle 3 \nabla \tilde{B} - 4 \nabla \tilde{U} \, , \, \nabla \tilde{A} \rangle 
= - \tilde{\psi}^2 \cdot (\tilde{\Omega} - \tilde{A}) \, ,
\end{equation}
$\tilde{\psi}^2$ defined like $\psi^2,$ (\ref{psi2}), but with $5$-{\em lifted} functions 
(on $\R^5).$

\subsection*{B. Regularity and boundedness of the metric functions}

Let us see how conditions (\ref{cond:C1C2})-(\ref{cond:bound0}) transmit through
the $5$-lift. 
First, from conditions (\ref{cond:C1C2}) and (\ref{cond:axisym}) it follows
\begin{equation}
\label{cond:C1_n}
\tilde{K}, \tilde{U}, \tilde{B}, \tilde{A} \, \in \, C^1(\R^5) \, ,
\end{equation}
because (for $v = K, U, B, A)$ \ $v \in C^1(\RR)$ \ and \ $\partial_{\rho} v \to 0$ 
as $\rho \to 0$ \ imply \ $\tilde{v} \in C^1(\R^5).$ 

\vspace{4mm}

\noindent
{\em Note}\/: In fact, as seen in
Ref.~6, with the use of these mathematical tools $(n$-lift and
meridional cut, for different numbers $n),$ the elliptic
system of field equations (\ref{eq:Einstein}) may be regarded as a set of
{\em Poisson-like equations}, where the nonlinearities (quadratic terms in
the first derivatives of the metric functions) are contained in the 
inhomogeneous terms on the right hand side. Making the weak requirement
that the metric functions and their derivatives are essentially bounded,
$\tilde{v}, \, \nabla \tilde{v} \in L^{\infty},$ since
also $\tilde{\varepsilon}, \, \tilde{p} \in L^{\infty}$ (by condition
(\ref{cond:cont_p}) and $\tilde{\varepsilon} = \tilde{p} = 0$ in the exterior), and 
$\tilde{\Omega} \in L^{\infty}$ (by (\ref{cond:Om_C1})), we have that 
the right hand side is essentially bounded. 
Then, by the regularity of Poisson's integral \cite{fn2}, 
(at least) $\tilde{v} \in C^{1,\alpha}$ 
for some $\alpha < 1;$ in particular, $\tilde{v} \in C^{1},$ i.e.~(\ref{cond:C1_n}).
This justifies requirements (\ref{cond:C1C2}) and (\ref{cond:axisym}) in Sec.~II.

\vspace{4mm}

Combining (\ref{cond:C1C2}) and (\ref{cond:axisym}) we obtain also
that the $5$-lifted metric functions are class $C^2$ in the
interior and in the exterior of the star (in $\R^5).$ That is, denoting

\begin{equation}
\label{def:calI}
{\cal I}  :=  \{
(x_1,\ldots,x_5)\in\R^5 \; | \,
   ((x_1^{\ 2} + \cdots + x_{4}^{\ 2})^{1/2}\, , \, x_5)\in I \} 
         \subset \R^5 
\end{equation}
and, analogously, ${\cal E}$ and ${\cal S},$ from $E$ and $S$ (cf.~(\ref{def:EIS})),
respectively, we have for $\tilde{v} = \tilde{K}, \tilde{U}, \tilde{B}, \tilde{A}$
\begin{equation}
\label{cond:C1C2_v}
\tilde{v} \, \in \, 
C^2({\cal I}) \cap C^2({\cal E}) \cap C^1(\R^5) 
\, .
\end{equation}
The asymptotic flatness condition (\ref{cond:as-flat})
implies, through the $5$-lift, that
\begin{equation}
\label{cond:as-flat_v}
\mbox{as \ {\it \small R}} = \|x\| = ({x_1}^{\ 2} +\ldots+ x_5^{\ 2})^{1/2}\to\infty, \qquad 
\tilde{v} \to 0 
\quad \mbox{and} \quad
\nabla \tilde{v} \to {\mathbf 0} 
\, ;
\end{equation}
but $\tilde{v} \in C^1(\R^5),$ that is, $\tilde{v} \in C^0(\R^5)$ and
$\nabla \tilde{v} \in \left[ C^0(\R^5) \right]^5,$ yielding, together with
conditions (\ref{cond:as-flat_v}), their respective boundedness,
\begin{eqnarray}
\label{cond:bound0_v}
|\tilde{v}| < \infty \quad \mbox{and} \quad
\|\nabla \tilde{v}\| < \infty \ \ \mbox{in} \ \R^5 \, .
\end{eqnarray}

\subsection*{C. Notation convention and roundup}

We have seen in Subsec.~III.A that Eq.~(\ref{eq:A-eq}) is equivalent to Eq.~(\ref{eq:A-eq0}) 
through the $5$-lift and the meridional cut, (\ref{def:n-lift}) and (\ref{def:merid_cut}) 
for $n=5.$
Furthermore, the
$5$-lift leaves regularity and boundedness properties invariant; see Subsec.~III.B.

\vspace{2mm}

\noindent
{\it Convention}\/:
For simplicity in the notation, we omit throughout the following the symbols
`$\;\tilde{}\;$' for all $5$-{\em lifted} functions we use.
(Once it has been seen how regularity and boundedness properties transmit
from the functions defined on $\RR$ to the lifted ones (on $\R^5),$ and
since they are equivalent in terms of positivity, and no explicit reference
to the first ones will appear throughout the following section,
this notation convention seems appropriate.)

\vspace{1mm}

Accordingly, we write Eq.~(\ref{eq:A-eq}) in the form

\vspace{2mm}

\fbox{
\parbox{12.5cm}{
\begin{eqnarray*}
L_0 \, A & = & - \psi^2 \cdot (\Omega - A) \, ,\\
& & \\
\mbox{with} & & L_0 \, A := \Delta A +
\langle 3 \nabla B - 4 \nabla U \, , \, \nabla A \rangle \, , \\
& & \\
& & \psi^2 := 16 \pi \frac{e^{2K}}{N} (\varepsilon + p) \ \ge 0 
\ \ (= 0 \ \, \mbox{in} \ {\cal E} ) \, , \\
& & \mbox{and} \quad N := e^{2U} - \rho^2 e^{2(B-U)}(\Omega - A)^2 
\ \ge \ \mbox{const.} > 0 \ \, \mbox{in} \ \ol{\cal I}
\, ,
\end{eqnarray*}
}}
\hfill
\parbox{1cm}{\begin{eqnarray}
\label{eqL0} \\
& & \nonumber \\
& & \label{L0} \\
& & \nonumber \\
& & \label{ps2} \\
& & \nonumber \\
& & \nonumber
\end{eqnarray}}

\vspace{2mm}

\noindent where $K, U, B, A : \R^5 \rightarrow \R,$ axisymmetric around the
$x_5$-axis. (Notice, Eq.~(\ref{eqL0}) is so defined in the whole spacetime,
interior (fluid) and exterior (vacuum), $\ol{\cal I} \cup {\cal E} = \R^5,$ 
but in the exterior $\psi^2 \equiv 0$ $(\varepsilon = p = 0)$ and the vacuum
field equation is recovered, $L_0 \, A = 0$ in ${\cal E}.)$ \ Also, we have 
(\ref{cond:C1C2_v})-(\ref{cond:bound0_v}), i.e.~(with the notation convention)
\begin{equation}
\label{cond:C1C2_n}
K, \, U, \, B, \, A \, \in \, 
C^2({\cal I}) \cap C^2({\cal E}) \cap C^1(\R^5) \, ,
\end{equation}

\begin{eqnarray}
\mbox{as \ {\it \small R}} = \|x\| = ({x_1}^{\ 2} +\ldots+ x_5^{\ 2})^{1/2}\to\infty, \quad 
& & K, \,  U, \, B, \, A \to 0 \label{cond:as-flat_n} \, , \\
& & \nabla K, \; \nabla U, \; \nabla B, \; \nabla A \, \to {\mathbf 0} 
\label{cond:as-flat_D_n}
\, .
\end{eqnarray}

\begin{eqnarray}
& & |K|, \; |U|, \; |B|, \; |A| \; < \infty \label{cond:bound0_n}\\
\mbox{and} & & \|\nabla K\|, \; \|\nabla U\|, \; \| \nabla B\|, \; \|\nabla A\| \;
< \infty \ \ \mbox{in} \ \R^5 \label{cond:bound0_D_n}
\, .
\end{eqnarray}

\bigskip

\noindent Equation~(\ref{eqL0}), i.e.
\begin{equation}
\label{eqL}
L \, A := L_0 \, A - \psi^2 \cdot A = - \psi^2 \cdot \Omega \, ,
\end{equation}
writes then
$$
L \, A \, \equiv \, a_{ij}(x) \, \partial_{ij} A 
\, + \, b_i(x) \, \partial_{i} A \, +  \, c(x) \, A \, = \, g(x) 
$$
\begin{equation}
\label{def:cf}
\parbox{11cm}{
\parbox{10cm}{
\begin{eqnarray*}
\mbox{with} \qquad a_{ij} & \equiv & \mbox{const.} \ = \delta_{ij} 
\ \ (= 1 \ \mbox{if} \ i=j, \ \mbox{and} \ = 0 \ \mbox{otherwise}) \, , \\
b_i & = & 3 \, \partial_i B - 4 \, \partial_i U \ \ \ \ \
(\forall \, i,j \in \{1,\ldots, 5\}), \ \ \mbox{and} \\ 
c & = & - \psi^2 \ \ (\le 0) \, , \\
g & = & c \, \Omega \, ,
\end{eqnarray*}
}}
\end{equation}
(where repeated indices indicate summation from $1$ to $5)$. 
The flat $5$-dimensional Laplacian $\Delta,$ in (\ref{L0}),
$(a_{ij} \equiv \delta_{ij}),$ and hence $L,$ is obviously strictly and uniformly 
elliptic everywhere. The coefficients $b_i$ are measurable and bounded functions everywhere, 
because $B$ and $U$ are $C^1$, (\ref{cond:C1C2_n}), and have bounded derivatives, 
(\ref{cond:bound0_D_n}). On the other hand, for the coefficient $c$ 
(cf.~(\ref{ps2})), since 
(i) the metric functions are 
continuous, (\ref{cond:C1C2_n}), and bounded, (\ref{cond:bound0_n});
(ii) $p$ is continuous everywhere, (\ref{cond:cont_p}), and has compact support;
(iii) $\varepsilon$ is continuous in the (closed) interior $\ol{\cal I},$ 
from (\ref{cond:cont_p});
(iv) $\Omega$ is in particular continuous (in $\ol{\cal I}),$
\ (\ref{cond:Om_C1}), and, hence, measurable; and 
(v) $N \ge$ const. $>0$ (also in $\ol{\cal I}),$ \ (\ref{N>0});
it follows that $c \equiv -\psi^2$ is measurable and bounded in the interior ${\cal I},$
and, since $\psi^2 \equiv 0$ $(\varepsilon = p = 0)$ in the exterior ${\cal E},$ and the boundary 
(the star's surface) $\partial {\cal I} = {\cal S}$ 
is a set of measure zero, we have that the coefficient $c$ is 
measurable and bounded everywhere. This will allow us in the following section 
to apply maximum principles in the classical and in the generalized sense to the 
operator $L$ (and $L_0);$ see Appendixes A and B.

\section*{\normalsize IV. BOUNDS ON THE DRAGGING RATE}

\subsection*{A. Positivity of the dragging rate}

\begin{prop} \hfill
\\
\label{prop:A0}
If the distribution of angular velocity of the fluid is non-negative (and non-trivial),
then the dragging rate is positive everywhere.
$$
\Omega \ge 0, \ \ \Omega \not \equiv 0 \ \ \
\Longrightarrow \ \
A > 0 \, .
$$
\end{prop}

\proof
Consider the domain $G$ defined by a ball in $\R^5$ centered at the origin
$x = {\bf 0}$ and of arbitrarily large radius $\sigma,$ \ 
\begin{equation}
G := {\cal B}_ {\sigma} ({\bf 0}) \, \subset \R^5 \, .
\label{def:G}
\end{equation}
Since $A$ is continuously differentiable in $\R^5,$ cf.~(\ref{cond:C1C2_n}),
so is in particular in $G;$ but $A$ and $\nabla A$ continuous in $\R^5$ implies that 
they are $2$-integrable (are in $L^2)$ in $G;$ consequently,
\begin{equation}
\label{cond:on_A}
A \, \in \, W^{1,2}(G) \cap C^1(G)
\, .
\end{equation}
Hence, the strictly elliptic linear partial differential equation (in $A)$ with 
measurable and bounded coefficients, (\ref{eqL}), is 
satisfied in a generalized sense in $G;$ see Appendix~B.
Remarkably, whenever $\Omega \ge 0,$ Eq.~(\ref{eqL}) 
yields the differential inequality 
\begin{equation}
\label{eq:deA}
L \, A \le 0 \quad \mbox{in} \ G
\, ,
\end{equation}
i.e.~$A$ is a generalized supersolution relative to the operator $L$ 
and the domain $G.$ We pay now special attention to the behavior of $A$ on the boundary: 
since the radius of the ball $G,$ $\sigma,$ is arbitrary, 
we can make it sufficiently large $(\sigma \to \infty)$ such that, by the asymptotic
flatness condition on $A$ (cf.~(\ref{cond:as-flat_n})), $A$ is arbitrarily small on 
$\partial G,$
\begin{equation}
\label{A_as_flat}
\lim_{\sigma \to \infty} \left. A \right|_{\partial G} = 0 \, .
\end{equation}
We first observe that $A \not\equiv$ const. (because, by (\ref{A_as_flat}) and $A \in C^0(\R^5),$ 
would be $A \equiv$ const.$\, = 0,$ which yields, by Eq.~(\ref{eqL}), $\Omega \equiv 0,$
and we are assuming $\Omega \not\equiv 0).$ Hence, by the strong minimum principle,
Theorem~\ref{thm:stMPg} in Appendix~B, applied to the differential inequality (\ref{eq:deA}),
$A$ cannot attain a non-positive minimum at an interior point of $G;$ using (\ref{A_as_flat}), 
we conclude then $A > 0$ in $G,$ i.e.~everywhere.\qed

\begin{rem}{\em \label{remtoP1}\hskip-1ex.
A result analogous to Proposition~\ref{prop:A0} holds with the opposite sense of
the rotation; that is, if $\Omega \le 0$ $\, (\Omega \not \equiv 0),$ then $A<0.$
This follows because Eq.~(\ref{eqL0}) is invariant with respect to the simultaneous
change of sign $(\Omega,A) \rightarrow (-\Omega,-A).$
}
\end{rem}

\subsection*{B. Upper bound $\Omega$. Positivity of the angular momentum density}

Hereafter we discuss the sign of the difference $\, \Omega - A.$ Remarkably,
this determines the sign of \ $u^t u_{\phi},$ which, with assumption (\ref{eq:energy-cond}),
is the sign of the {\em angular momentum density},
integrand of the total angular momentum, given by the ``volume'' integral\cite{H-Sh}
$\, J = \int_{\cal I} 2 \pi \, T_{\phi}^{\; t} \, (-{\rm g})^{\! 1/2} d x,$ where 
$\, {\rm g} \equiv \det({\mathbf g}) \;$ and $\; T_{\phi}^{\; t} = (\varepsilon + p) \, u^t u_{\phi}.$

\subsubsection*{1. In the rigidly rotating case}

\begin{prop} \hfill
\\
\label{prop:upb_rr}
In the particular case of rigid rotation, 
with \ $\Omega \equiv$ {\rm const.}$\, =: \Omega_{\ast} > 0,$ 
$$
0 < A < \Omega_{\ast}
$$
holds everywhere. As a consequence, in this case, $u^t u_{\phi},$ and, hence, 
the angular momentum density, is non-negative.
\end{prop}

\proof
We consider Eq.~(\ref{eqL0}) for $\Omega \equiv$ const.$\, = \Omega_{\ast} > 0,$ \ 
i.e. \ $L_0 \, A  = - \psi^2 \cdot (\Omega_{\ast} - A),$ which,
since the differential operator $L_0,$ (\ref{L0}), is free from linear term,
can be rewritten in the form
$$
L_0 (A - \Omega_{\ast})  = - \psi^2 \cdot (\Omega_{\ast} - A) 
\, ,
$$
or, denoting again the differential operator $L := L_0 - \psi^2$ and defining 
\begin{equation}
\label{def:w}
w(x) := A(x) - \Omega_{\ast}
\end{equation}
in the whole spacetime, $x \in \ol{\cal I} \cup {\cal E} = \R^5$ 
(as already $5$-lifted function; cf.~Subsec.~III.A),
$$
Lw = L (A - \Omega_{\ast}) = L_0 (A - \Omega_{\ast}) - \psi^2 \cdot (A - \Omega_{\ast}) = 0 
\quad \mbox{in} \ \R^5
\, .
$$
We have then the strictly elliptic linear (in $w)$ equation
\begin{equation}
\label{eq:w}
Lw = 0 \qquad \ \ 
\mbox{(in particular) in} \ G \equiv {\cal B}_ {\sigma} ({\bf 0}) \, \subset \R^5
\, ,
\end{equation}
where the radius $\sigma$ is arbitrary, with
$w \in W^{1,2}(G) \cap C^1(G),$ by (\ref{cond:on_A}) and (\ref{def:w}).
On the other hand, by the condition of 
asymptotic flatness on $A$ (in~(\ref{cond:as-flat_n})), $A$ is arbitrarily small on 
$\partial G,$ provided that $\sigma$ is sufficiently large, i.e.~(\ref{A_as_flat}); consequently,
\begin{equation}
\label{cond:w_bound}
\lim_{\sigma \to \infty} \left. w \right|_{\partial G} = - \Omega_{\ast} < 0 \, .
\end{equation}
Since $w \not\equiv$ const. (because, by (\ref{cond:w_bound}) and continuity, would be
$w \equiv$ const.$\, = -\Omega_{\ast}$ in $G,$ that is, $A \equiv$ const.$\, = 0$ in $G,$
which is not allowed, by Eq.~(\ref{eqL}), since here $\Omega \equiv \Omega_{\ast} > 0),$
applying the strong maximum principle,
Theorem~\ref{thm:stMPg} in Appendix~B, to Eq.~(\ref{eq:w}), we get that
$w$ cannot attain a non-negative maximum at an interior point of $G;$ hence, using (\ref{cond:w_bound}), 
$w < 0$ in $G$ (everywhere), i.e.~$A < \Omega_{\ast}$ everywhere.
Moreover, $A > 0$ everywhere, by Proposition~\ref{prop:A0}. This establishes the 
conclusion of the proposition. \qed

\vspace{2mm}

\noindent Observe, in the static case, $\Omega_{\ast} = 0,$ we would have 
$Lw = 0$ and $\lim_{\sigma \to \infty} \left. w \right|_{\partial G} = 0,$ following,
by the strong maximum and minimum principles, $w \equiv 0,$ i.e. 
$A \equiv \Omega_{\ast} = 0;$ as expected, $A \equiv 0.$

\begin{rem}{\em \label{remtoP2}\hskip-1ex.
Likewise, if $\Omega \equiv$ const.$\, \equiv \Omega_{\ast} < 0,$ then
$0 > A > \Omega_{\ast}$ everywhere, and, hence, the angular momentum density is
non-positive. We obtain this by applying Proposition~\ref{prop:upb_rr}
to the function $\hat{A} := -A,$ solution of Eq.~(\ref{eqL0}) for 
$\hat{\Omega}_{\ast} := - \Omega_{\ast} > 0$ \ (cf.~Remark~\ref{remtoP1}).
More explicitly, the angular momentum density of a rigidly rotating stellar model has the same
sign as the angular velocity of the fluid.
Also, as a result, we have for a fluid rotating rigidly with  
$\Omega \equiv$ const.$\, \equiv \Omega_{\ast} \neq 0$ 
$$
0 < |A| < |\Omega_{\ast}| \, .
$$
}
\end{rem}

\subsubsection*{2. In the general (differentially rotating) case}

In the following we shall assume that a function ${\cal F}$ 
(to be specified) has been given,
and we have a solution of the problem, that is, (four) metric functions, $K,U,B,$ and $A,$
and a fluid angular velocity distribution, $\Omega,$ satisfying the (four) field equations 
(\ref{eq:Einstein}) (in particular, the elliptic equation for $A,$ Eq.~(\ref{eqL0})) and 
Eq.~(\ref{cond:dr}), $u^t u_{\phi} = {\cal F}(\Omega).$
(Notice, in the interior, where the matter terms do not vanish $(p > 0,$ $\varepsilon \ge 0),$
substituting into the equation of motion (\ref{eq:Euler}) (with (\ref{def:acc}))
its integrability condition, 
i.e.~Eq.~(\ref{cond:dr}), and the equation of state, Eq.~(\ref{eq:eos}), we obtain the pressure,
$p,$ and the energy density, $\varepsilon,$ as functions of $\rho, U, B, A,$ and $\Omega.)$

\vspace{1mm}

\noindent Remarkably, $u^t u_{\phi}$ may be written
\begin{eqnarray}
u^t u_{\phi} & \equiv & 
\frac{\rho^2 e^{2(B-U)} (\Omega - A)}{e^{2U} - \rho^2 e^{2(B-U)}(\Omega - A)^2}  =  \nonumber \\
& & \frac{\varrho^2 (\Omega - A)}{1 - \varrho^2 (\Omega - A)^2}  =: 
\Phi(\varrho \, , \, \Omega - A) 
\qquad  \mbox{with} \ \ \varrho := \rho \, e^{B-2U} \label{def:Phi} 
\, , 
\end{eqnarray}
where, from (\ref{N>0}), \ $1 - \varrho^2 (\Omega - A)^2 = N e^{-2U} \ge$ const. $> 0$ 
in $\ol{\cal I}.$ \ With the defined function (\ref{def:Phi}), Eq.~(\ref{cond:dr}) writes
\begin{equation}
\label{dr}
\Phi(\varrho \, , \, \Omega - A) = {\cal F} (\Omega) \, .
\end{equation}

\noindent {\bf Lemma}\\
{\it Assume 
\begin{enumerate}
\item[{\em i.}] \quad the function \ ${\cal F}: \R \longrightarrow \R$ \ is {\em strictly decreasing}, and
 
\item[{\em ii.}] \quad $\exists$ a constant $\Omega_c$ \ $(|\Omega_c| < \infty)$ \ such that
${\cal F} (\Omega_c) = 0,$ 
\end{enumerate}
then, at each interior point (in $\ol{\cal I}$), where Eq.~{\em(\ref{dr})}, $\Phi = {\cal F},$ 
is satisfied, the following holds
\\
\parbox[c]{15cm}{
\begin{eqnarray*}
A < \Omega & \iff & A < \Omega \le \Omega_c \quad (A < \Omega_c)\\
A > \Omega & \iff & A > \Omega \ge \Omega_c \quad (A > \Omega_c)\\
A = \Omega & \iff & A = \Omega = \Omega_c \quad (A = \Omega_c) \, .
\end{eqnarray*}
}
\hfill
\parbox{0.5cm}{\begin{equation}\label{p-res}\end{equation}}
\\
}
\noindent
{\em Note~1}\/:
Due to (i), $\Omega_c$ (defined in (ii)) is unique.
Also, observe, $\Omega_c$ exists and coincides with the (constant) value of 
$\Omega$ on the rotation axis, provided that 
part of the axis, $\varrho = 0,$ is in the interior,
${\cal I},$ (i.e. if the rotating fluid does not have toroidal topology). 
This is because
at points in  $\{ \varrho = 0 \} \cap {\cal I} \neq \emptyset,$ since 
$\left. \Phi \right|_{\varrho = 0} = 0$ and $\Phi = {\cal F}$ in ${\cal I},$
we have $\left. {\cal F}(\Omega) \right|_{\varrho = 0} = 0;$ and ${\cal F}$ is,
by requirement (i), invertible; yielding $\left. \Omega \right|_{\varrho = 0} =$ 
const.$\, = \Omega_c.$ 

\vspace{2mm}

\noindent
{\em Note~2}\/:
Observe, if ${\cal F} \in C^1$ and ${\cal F}\,' < 0,$ then, since 
$\partial_{\Omega} \Phi \ge 0,$ Eq.~(\ref{dr}) can be solved for $\Omega,$
by virtue of the implicit function theorem, yielding $\Omega = \Omega(\rho, U, B, A);$
and, by the regularity of the metric functions, (\ref{cond:C1C2_n}), it follows
in particular $\Omega \in C^1 \left(\ol{\cal I} \right),$ requirement (\ref{cond:Om_C1}).

\vspace{2mm}

\noindent
{\em Note~3}\/:
It should be stressed that, since $\Phi$ is an {\em increasing} function in $\Omega,$
choosing the function ${\cal F}$ strictly {\em decreasing} (requirement (i)),
Eq.~(\ref{dr}) has a unique solution in $\Omega$ (``curve'' solution with $\varrho$
variable). Specially, this makes likely the existence of functions
$\Omega,$ $K,$ $U,$ $B,$ and $A,$ solutions of the field equations and Eq.~(\ref{cond:dr}).
Indeed, in numerical works concerning differential rotation the ansatz for the ${\cal F}$-law
${\cal F}(\Omega) = R_0^{\, 2} (\Omega_c - \Omega),$ where $R_0$ is a free parameter
describing the length of scale over which $\Omega$ changes, is
generally used, and they claim they have a solution. (See~e.g.~Refs.~15-17.)

\vspace{4mm}

\proof
We consider a point $x \in \ol{\cal I}$ where the metric functions and the fluid
angular velocity are solution, in particular, with reference to Eq.~(\ref{dr}),
the functions $\Phi$ and ${\cal F}$ valued at this point ``intersect'' each other, i.e.
$$
\Phi(\varrho(x) \, , \, \Omega(x) - A(x)) = {\cal F} (\Omega(x)) 
\qquad ( \forall x \in \ol{\cal I} )
\, .
$$
From requirements (i) and (ii), it follows (at each interior point) 
sign$({\cal F})$ $=$ sign$(\Omega_c - \Omega).$ 
As regards $\Phi$ (at the interior point), 
on the axis $(\varrho=0)$ it obviously vanishes, cf.~(\ref{def:Phi}); following,
from the relation $\Phi = {\cal F}$ (in $\ol{\cal I}),$  $\Phi = {\cal F} = 0$ and, thus,
$\Omega = \Omega_c$ on the axis. Outside the axis $(\varrho \neq 0)$ we have 
sign$(\Phi)$ $=$ sign$(\Omega - A),$ and, hence, by $\Phi = {\cal F}$ (in $\ol{\cal I}),$ 
sign$(\Omega_c - \Omega)$ $=$ sign$(\Omega - A)$ outside the axis. This yields (\ref{p-res}).
\qed

\vspace{1mm}

We are now in a position to get one of the main results of this work
in the general differentially rotating case, namely,
\begin{prop} \hfill
\\
\label{prop:upb_dr}
If the ${\cal F}$-law (in Eq.~{\em (\ref{cond:dr})} ) specifying the model of 
differential rotation is chosen such that 
\begin{enumerate}
\item[{\em i.}] \quad ${\cal F}: \R \longrightarrow \R$ \quad strictly decreasing, 

\item[{\em ii.}] \quad $\exists$ a constant $\Omega_c$ \ $(|\Omega_c| < \infty)$ such that 
${\cal F} (\Omega_c) = 0,$ and

\item[{\em iii.}] \quad $\Omega_c > 0,$
\end{enumerate}
then 
\begin{equation}
\label{res_dr}
0 < A < \Omega \le \Omega_c \quad \mbox{in} \ \ol{\cal I} 
\, ;
\end{equation}
in particular, $u^t u_{\phi} \ge 0,$ and,
hence, the angular momentum density is non-negative. Moreover, 
\begin{equation}
\label{res_drA}
0 < A < \max_{\cal S} \Omega \le \Omega_c \quad \mbox{in} \ {\cal E}
\, .
\end{equation}

\end{prop}
\noindent
{\em Note}\/: 
As remarked above, if the interior (fluid) contains points of the axis,
then condition (ii) is already guaranteed, and $\Omega_c$ is the constant value of 
$\Omega$ on the axis; cf.~Note~1 in the previous lemma. See also Notes~2~and~3.
And observe, requirement (iii) is in principle much 
weaker than $\Omega > 0,$ but, as seen in the conclusion of this proposition,
$\Omega > 0$ already follows. Furthermore, 
the fact that $\Omega \le \Omega_c$ in $\ol{\cal I}$ shows that in differentially 
rotating stars the core may rotate faster than the envelope, so that the core can be
supported by rapid rotation before mass shedding is reached at the equator \cite{BSS}.

\vspace{2mm}

\proof
We divide the proof in four steps.

\begin{description}

\item[1st step:] Let us see first \ $A \le \Omega$ \ in $\ol{\cal I}.$

\vspace{1mm}

\noindent Suppose (to get a contradiction) $A(x_0) > \Omega(x_0)$ for some 
$x_0 \in \ol{\cal I}.$
We have seen, in the previous lemma, cf.~(\ref{p-res}), that this is equivalent to
$A(x_0) > \Omega(x_0) \ge \Omega_c$; and, hence, using hypothesis (iii),
$A(x_0) > \Omega_c > 0.$ Therefore, by the continuity of $A$ (indeed $A \in C^1,$
cf.~(\ref{cond:C1C2_n})) and the asymptotic flatness
$(\lim_{\|x\| \to\infty} A = 0,$ \ cf.~(\ref{cond:as-flat_n})),
we infer that 
\\
\noindent
\parbox[c]{13cm}{
\vbox{
\hbox{}\hfill
\hbox{
$\exists \;$ an open and connected neighborhood of $\, x_0, \,$ ${\cal N}_{x_0} \subset \R^5, \,$ such that 
}
\hfill\hbox{}
}
\vspace*{-4.5ex}
\begin{eqnarray*}
& A > \Omega_c & \ \ \mbox{in} \ {\cal N}_{x_0} \\
\mbox{and} & A = \Omega_c & \ \ \mbox{on} \ \partial {\cal N}_{x_0}
\, .
\end{eqnarray*}
}
\hfill
\parbox[b]{0.5cm}{\begin{equation}\label{neig}\end{equation}}
\\
We distinguish two cases:

\vspace{1mm}

\noindent {\bf case 1:} \ ${\cal N}_{x_0} \cap {\cal E} = \emptyset,$ that is, 
the neighborhood is contained in the interior, ${\cal N}_{x_0} \subseteq {\cal I}.$\\
Thus we have, again using the previous lemma, 
$A > \Omega \ge \Omega_c > 0$ in ${\cal N}_{x_0};$
particularly, $\Omega - A < 0$ in ${\cal N}_{x_0},$ and, 
therefore, by Eq.~(\ref{eqL0}),
$$
L_0 \, A > 0 \qquad \mbox{in} \ {\cal N}_{x_0} \subseteq {\cal I}
\, .
$$
From (\ref{cond:C1C2_n}), in particular, $A \in C^2({\cal I}) \cap C^0( \ol{\cal I}),$
and, hence, $A \in C^2({\cal N}_{x_0}) \cap C^0(\ol{\cal N}_{x_0}).$ And applying 
the weak maximum principle, Theorem~\ref{thm:wMP_L0} in Appendix~A, to the operator
$L_0$ (on $A)$ in ${\cal N}_{x_0},$ we obtain that the maximum of $A$ is reached on
the boundary, i.e.
$$
\max_{\ol{\cal N}_{x_0}} A = \max_{\partial {\cal N}_{x_0}} A \, ,
$$
contradicting (\ref{neig}).

\vspace{1mm}

\noindent {\bf case 2:} ${\cal N}_{x_0} \cap {\cal E} \neq \emptyset.$ (Notice, here is
included the case where $x_0 \in \partial {\cal I} = {\cal S}.)$  We denote
\begin{eqnarray*}
{\cal I}_{x_0} & \equiv & {\cal N}_{x_0} \cap {\cal I} \quad \subseteq {\cal I} \\
{\cal E}_{x_0} & \equiv & {\cal N}_{x_0} \cap {\cal E} \quad \subset {\cal E} \\
& & \\
\mbox{and} \ \ \Gamma & \equiv & {\cal N}_{x_0} \cap {\cal S} \, .
\end{eqnarray*}
Observe, $\Gamma \neq \emptyset,$ because ${\cal N}_{x_0}$ is connected, 
$x_0 \in {\cal N}_{x_0} \cap \ol{\cal I} \neq \emptyset,$
and, by assumption in this case, ${\cal N}_{x_0} \cap {\cal E} \neq \emptyset.$ Moreover,
$\Gamma = \ol{\cal I}_{x_0} \cap \ol{\cal E}_{x_0} =
\partial {\cal I}_{x_0} \cap \partial {\cal E}_{x_0}.$ 

\noindent Thus, we have
\begin{description}
\item[in the interior,] from (\ref{neig}),
\begin{eqnarray*}
& A > \Omega_c & \ \ \mbox{in} \ \, {\cal I}_{x_0} \cup \Gamma \quad \subset \, {\cal N}_{x_0}
\qquad (\Gamma \subset \partial {\cal I}_{x_0})\\
& A = \Omega_c & \ \ \mbox{on} \ \, \partial {\cal I}_{x_0} \setminus \Gamma \quad
\subset \, \partial {\cal N}_{x_0}
\, ;
\end{eqnarray*}
and, applying the weak maximum principle (Theorem~\ref{thm:wMP_L0} in Appendix~A)
to the differential inequality (cf.~(\ref{eqL0}))
$$
L_0 \, A > 0 \qquad \mbox{in} \ {\cal I}_{x_0} \subset {\cal I} 
$$
(again using (\ref{p-res}), $A > \Omega \ge \Omega_c$ in ${\cal I}_{x_0}),$
with $A \in C^2({\cal I}_{x_0}) \cap C^0(\ol{\cal I}_{x_0}),$ 
it follows
$$
\max_{\ol{\cal I}_{x_0}} A = \max_{\partial {\cal I}_{x_0}} A = \max_{\Gamma} A 
\, ;
$$

\item[in the exterior,] we have analogously, from (\ref{neig}),
\begin{eqnarray*}
& A > \Omega_c & \ \ \mbox{in} \ \,  {\cal E}_{x_0} \cup \Gamma \quad \subset \, {\cal N}_{x_0}
\qquad (\Gamma \subset \partial {\cal E}_{x_0})\\
& A = \Omega_c & \ \ \mbox{on} \ \, \partial {\cal E}_{x_0} \setminus \Gamma \quad
\subset \, \partial {\cal N}_{x_0} 
\, .
\end{eqnarray*}
(Note, `the point $\infty$' is not included in $\partial {\cal E}_{x_0},$ 
because $\Omega_c > 0$ and $A$ is asymptotically flat, $\left. A \right|_{\infty} = 0.)$
But, in the exterior, ${\cal E},$
$\psi^2 \equiv 0,$ and we have the elliptic equation for $A$ 
$$
L_0 \, A = 0 \qquad \mbox{in} \ {\cal E}_{x_0} \subset {\cal E}
\, ;
$$
as a consequence, again by virtue of the maximum principle 
(Theorem~\ref{thm:wMP_L0} in Appendix~A) now in ${\cal E}_{x_0}$
$$
\max_{\ol{\cal E}_{x_0}} A = \max_{\partial {\cal E}_{x_0}} A = \max_{\Gamma} A
\, .
$$
\end{description}
We therefore have 
$$
\max_{\ol{\cal I}_{x_0}} A = \max_{\ol{\cal E}_{x_0}} A = \max_{\Gamma} A =: A(x_1),
\quad \mbox{for some} \ x_1 \in \Gamma
\, .
$$
Thus, $\left( \, \ol{\cal N}_{x_0} = \ol{\cal I}_{x_0} \cup \ol{\cal E}_{x_0} \right)$
$$
\max_{\ol{\cal N}_{x_0}} A = A(x_1), \quad \mbox{for some} \ x_1 \in \Gamma \subset {\cal N}_{x_0}
\quad \mbox{($x_1$ interior point)}
\, ;
$$
and, since $A \in C^1(\R^5),$ in particular, 
$A \in C^1({\cal N}_{x_0}),$ it follows 
\begin{equation}
\label{nabla0}
\left. \nabla A \right|_{x_1} = {\mathbf 0}
\, .
\end{equation}
However, this is not possible, because, on the other hand, $x_1$ is a point of the star's surface
$x_1 \in \Gamma \subset {\cal S},$ and, from the assumptions on the stellar model,
$\partial {\cal I} = {\cal S} $ satisfies an exterior sphere condition everywhere, i.e.~$\partial {\cal E} 
= {\cal S} \cup \{ \infty \}$ satisfies at each point of ${\cal S}$ (in particular, 
at $x_1)$ an interior sphere condition (cf.~Definition in Appendix~A). This allows us to apply
the called boundary-point lemma, Theorem~\ref{thm:bound-point} in Appendix~A,
for the operator $L_0$ in the exterior domain ${\cal E}_{x_0},$
with respect to the point $x_1 \in \Gamma \subset \partial {\cal E}_{x_0},$ being $A(x_1) = \max A \,$ in 
$\ol{\cal E}_{x_0}.$ \ And, since $A \not \equiv$ const. 
(because $A > \Omega_c > 0$ in ${\cal N}_{x_0},$ $A \in C^1(\R^5),$ 
and $\left. A \right|_{\infty} = 0),$ this yields
$$
\left.\langle {\boldsymbol \nu} , \nabla A \rangle \right|_{x_1} = 
\left. \partial_{{\boldsymbol \nu}} A \right|_{x_1} \neq 0, 
\qquad {\boldsymbol \nu} \equiv \ \mbox{outward pointing unit normal to} \ {\cal S} \ \mbox{at} \ x_1
\, ;
$$
contradicting (\ref{nabla0}). Consequently, 
\begin{equation}
\label{ineq1}
A \le \Omega \quad \mbox{in} \ \ol{\cal I}
\, .
\end{equation}

\item[2nd step:] \ $A < \Omega \le \Omega_c$ \ in $\ol{\cal I}.$

\vspace{1mm}

\noindent This can be seen as follows. From inequality (\ref{ineq1}) and using (\ref{p-res})
we also have
\begin{equation}
\label{ineq2}
\Omega \le \Omega_c \quad \mbox{in} \ \ol{\cal I}
\, ,
\end{equation}
and, combining (\ref{ineq1}) and (\ref{ineq2}),
\begin{equation}
\label{ineq3}
A \le \Omega_c \quad \mbox{in} \ \ol{\cal I}
\, .
\end{equation}
On the other hand, we have Eq.~(\ref{eqL0}), i.e.~ $L_0 \, A =  - \psi^2 \cdot (\Omega - A),$
satisfied everywhere, in particular, in the interior (in a classical sense). Let
$$
u(x) := A(x) - \Omega_c \qquad \forall x \in \ol{\cal I} \, .
$$
Since $\Omega_c$ is constant, we can rewrite Eq.~(\ref{eqL0}),
$$
L u := L_0 u - \psi^2 \cdot u = + \psi^2 \cdot (\Omega_c - \Omega) \ge 0 \quad \mbox{(by (\ref{ineq2}))}
\, .
$$
Hence, we have 
\begin{equation}
\label{Lu}
L u \ge 0 \quad \mbox{in} \ {\cal I}
\, ,
\end{equation}
where, like $A,$ \ $u \in C^2({\cal I}) \cap C^0(\ol{\cal I}),$ and 
\begin{equation}
\label{u<0}
u \le 0 \quad \mbox{in} \ \ol{\cal I}
\, ,
\end{equation}
by inequality (\ref{ineq3}). We want to see $u < 0.$ Suppose (to get a contradiction) that 
$u(\hat{x}) = 0$ for some $\hat{x} \in {\cal I};$ then, by (\ref{u<0}), 
$0 = u(\hat{x}) = \max_{\ol{\cal I}} u,$ $\, \hat{x} \in {\cal I}$ (interior point).
However, by the strong maximum principle, Theorem~\ref{thm:stMP_L} in Appendix~A, 
for (\ref{Lu}), $u$ cannot reach a non-negative maximum at an interior point
of ${\cal I},$ unless $u$ is a constant in ${\cal I}.$ That means, in our case, 
$u$ cannot vanish somewhere in ${\cal I}$ unless it vanishes identically in ${\cal I}.$
But $u \equiv$ const.$\, = u(\hat{x}) = 0$ in ${\cal I},$ i.e.
\begin{equation}
\label{Aconst}
A \equiv \, \mbox{const.} = \Omega_c > 0 \quad \mbox{in} \ {\cal I}, 
\qquad A \in C^1 \ \mbox{everywhere}
\, ,
\end{equation}
yields, in particular, 
\begin{equation}
\label{nabla01}
\nabla A = {\mathbf 0} \quad \mbox{on} \ \partial {\cal I} = {\cal S}
\, .
\end{equation}
On the other hand, in the exterior, ${\cal E},$ we have $L_0 \, A = 0,$ with
$A \in C^2({\cal E}) \cap C^0(\ol{\cal E}),$ and, by the weak maximum principle
(Theorem~\ref{thm:wMP_L0} in Appendix~A)
$$
\max_{\ol{\cal E}} A \; = 
\max_{\partial {\cal E} = {\cal S} \cup \{ \infty \}} A 
\, ,
$$
but, using asymptotic flatness $(\left. A \right|_{\infty} = 0)$ and (\ref{Aconst}),
actually,
$$
\max_{\ol{\cal E}} A = \max_{\cal S} A  =: A(x_1) \quad 
\mbox{for some} \ x_1 \in {\cal S} \subset {\partial {\cal E}}
\, .
$$
In particular, since $A \not \equiv$ const., the boundary-point lemma,
Theorem~\ref{thm:bound-point} in Appendix~A, applied to the operator 
$L_0$ in the exterior domain ${\cal E}$ (where, by assumption, an interior sphere condition
is satisfied in particular at $x_1 \in {\cal S} \subset \partial {\cal E})$ 
yields a non-vanishing outward normal derivative
$$
\left.\langle {\boldsymbol \nu} , \nabla A \rangle \right|_{x_1} = 
\left. \partial_{{\boldsymbol \nu}} A \right|_{x_1} \neq 0
\, ,
$$
in contradiction to (\ref{nabla01}). Therefore, $u < 0$ everywhere in ${\cal I},$
i.e.~$A < \Omega_c$ in ${\cal I};$ and, hence, also on $\partial {\cal I},$
because, by the weak minimum principle, Theorem~\ref{thm:wMP_L0} in Appendix~A,
applied to $L_0 \, A = - \psi^2 \cdot (\Omega - A) \le 0$ in ${\cal I}$ (by (\ref{ineq1})), 
we get $\min_{\ol{\cal I}} A = \min_{\partial {\cal I }} A.$ Therefore,
$A < \Omega_c$ in $\ol{\cal I},$ or, equivalently (cf.~(\ref{p-res})),
$$
A < \Omega \le \Omega_c \quad \mbox{in} \ \ol{\cal I}
\, .
$$

\item[3rd step:] \ $A > 0$ everywhere (i.e.~the same 
conclusion of Proposition~\ref{prop:A0}, but now using different hypotheses).

\vspace{1mm}

\noindent We have seen in the first step $A \le \Omega$ in $\ol{\cal I};$ 
which yields, $L_0 \, A \le 0$ in $\ol{\cal I}.$ On the other hand, $L_0 \, A = 0$ in ${\cal E}.$
Accordingly,
$$
L_0 \, A \le 0 \quad \mbox{everywhere in} \ \ol{\cal I} \cup {\cal E} = \R^5
\, .
$$
Applying now the strong minimum principle for generalized supersolutions, 
Theorem~\ref{thm:stMPg} in Appendix~B, and using asymptotic flatness, 
as was argued in the proof of Proposition~\ref{prop:A0},
it follows $A > 0$ everywhere. (Notice, here $A \not\equiv$ const., 
because, by asymptotic flatness and continuity, $A \equiv$ const. is equivalent to $A \equiv 0;$ 
by Eq.~(\ref{eqL}), also $\Omega \equiv 0,$ and, hence, from $\Omega - A \equiv 0,$ we would have 
(cf.~(\ref{def:Phi})) $\, 0 \equiv \Phi = {\cal F}(0);$ but this is not possible, since requirements 
(i) and (ii) imply ${\cal F} (0) > {\cal F} (\Omega_c) = 0.)$

Thus, $A > 0$ everywhere, in particular, in the interior; 
using now the result of the second step, we finally get (\ref{res_dr}),
$0 < A < \Omega \le \Omega_c$ in $\ol{\cal I}.$ 
Notice, hence, $\Omega > 0$ (in $\ol{\cal I}).$

\item[4th step:] \ $A < \max_{\cal S} \Omega \le \Omega_c$ in ${\cal E} \,.$

The elliptic equation holding in the exterior, 
$$
L_0 \, A = 0 \quad \mbox{in} \ {\cal E}
\, ,
$$
yields, by virtue of the weak maximum principle (Theorem~\ref{thm:wMP_L0} in Appendix~A), 
$$
\max_{\ol{\cal E}} A \; = \max_{\partial {\cal E} = {\cal S} \cup \{ \infty \}} A  
\, ;
$$
but, using asymptotic flatness $(\left. A \right|_{\infty} = 0),$ we actually have \
$\max_{{\cal S} \cup \{ \infty \}} A  = 
\max_{\cal S} A.$ On the other hand, we have seen $A < \Omega \le \Omega_c$ in particular in
$\partial {\cal I} = {\cal S},$ and ${\cal S}$ is compact. Hence,
$$
\mbox{in} \ {\cal E}, \quad 0 < A \le \max_{\ol{\cal E}} A 
= \max_{{\cal S} = \partial {\cal I}} A  < \max_{\cal S} \Omega \le \Omega_c
\, ,
$$
establishing also (\ref{res_drA}).
\qed

\end{description}

\begin{rem}{\em \label{remtoP3}\hskip-1ex.
Analogously as argued in Remarks~1~and~2, it is possible to ``reflect''
Proposition~\ref{prop:upb_dr}. Accordingly, in particular, 
in a model for a star which is rotating differentially with the function ${\cal F}$ either 
strictly decreasing with $\Omega_c > 0$ or strictly increasing with $\Omega_c < 0,$ \ 
${\cal F}(\Omega_c) = 0,$ the angular momentum density has the same sign as $\Omega_c,$
and, hence, as the angular velocity of the fluid.
Also, accordingly, the following holds
$$
0 < |A| < |\Omega| \le |\Omega_c| \quad \mbox{in} \ \ol{\cal I} 
\, ,
$$
$$
0 < |A| < \left|\max_{\cal S} \Omega\right| \le |\Omega_c| \quad \mbox{in} \ {\cal E}
\, .
$$
}
\end{rem}

\section*{\normalsize V. GENERAL BOUNDS. ROTATIONAL ENERGY}

\subsection*{A. Preliminary Observation}

Let $u:\R^n \to \R$ be a differentiable function, 
$\, V \! :\R^n \to \R^n$ be a vector field, and
$G\subset \R^n$ a domain where Gauss' theorem can be applied. 
Then, due to
$$
\D(u V) = \sum_i\partial_i(u V_i) =
\langle {\cal D} u, V \rangle + u \; \D V 
$$
(where $\langle . , . \rangle$ is the Euclidean scalar product, ${\cal D}$ is the 
gradient operator, and $\D V := \sum_i \partial_i(V_i),$ the divergence), we get,
integrating over $G$ and applying the Gauss theorem,
\begin{equation}
\label{gauss}
\int_G u \; \D V = -\int_G \langle {\cal D} u, V\rangle
+ \int_{\partial G} u \, \langle V, {\boldsymbol \nu} \rangle
\, ,
\end{equation}
where ${\boldsymbol \nu}$ is the outer unit normal of $\partial G$
\ (and, for simplicity in the notation, volume- and surface elements have
been dropped).

\subsection*{B. Appropriate form of the field equation}

The general (elliptic) field equation for $A,$ Eq.~(\ref{eq:A-eq0}), 
may be rewritten as follows:
\begin{equation}
\label{eq-div}
\D(\rho^3 e^{3B} e^{-4U}{\cal D} A) =
- f^2 \cdot (\Omega - A)
\, ,
\end{equation}
where ${\cal D} := (\partial_{\rho}, \partial_z)$ and  `$\D$' are 
the flat expressions in $\R^2,$ and
$$
f^2(\rho,z) \equiv f^2 := \rho^3 e^{3B - 4U} \psi^2 = 16\pi \, (\varepsilon + p) \, 
\frac{\rho^3 e^{3B} e^{2(K-2U)}}{e^{2U} - \rho^2 e^{2(B-U)}(\Omega - A)^2}
\, \ge 0
\, .
$$
Especially we have $f^2 \equiv 0$ in the exterior $E$ of the star (cf.~(\ref{def:EIS})).

Note, in this section (independent of Sec.~IV) we go back
to the field equation in the meridian plane coordinates, $(\rho,z) \in \RR,$ instead of
the $5$-{\em lifted} one (on $\R^5)$ (cf.~Subsec.~III.A).

\subsection*{C. Main Observation}

Multiplying Eq.~(\ref{eq-div}) by $A,$ and using Eq.~(\ref{gauss}),
by setting $u = A,$
$V=\rho^3 e^{3B}e^{-4U}{\cal D} A,$ and $G = \RR \subset \R^2$ 
(actually, we consider a ball in $\R^2$ centered at the origin 
of the coordinate system with arbitrarily large radius, ${\cal B}_ {\sigma}({\bf 0}) \subset \R^2,$
and take $G = {\cal B}_ {\sigma}({\bf 0})\cap(\RR) \subset \R^2,$ ${\sigma} \to \infty$), we obtain
$$
-\int_{I} f^2 A(\Omega-A) =
-\int_{\RR} \rho^3 e^{3B}e^{-4U} \|{\cal D} A \|^2 +
\int_{\partial(\RR)}
A\,\rho^3 e^{3B}e^{-4U} \langle {\cal D} A,\nu\rangle
\, ,
$$
where $I \subset \RR \subset \R^2$
represents the $(\rho,z)$-coordinates of the
interior of the star (note, $f^2$ vanishes in the exterior $E$). The first
term on the right hand side (which converges, since $\|{\cal D} A \|$ falls off
rapidly enough at the space-like ``infinity" \cite{fn4}) is obviously non-positive. 
And the second term 
vanishes, because of the asymptotic behavior of $A$ at spatial infinity\cite{fn4}, and 
because the integrand, due to the factor
$\rho^3,$ vanishes on the axis of rotation $\rho = 0,$ which is the other part of \ 
$\partial(\RR) = \{ \mbox{\it \small R} 
= (\rho^2+z^2)^{1/2} \to \infty \} \cup  \{\rho = 0 \}.$ 
Hence, we have found
\begin{equation}
\label{main}
\int_{I} f^2 \, A \, (\Omega-A) \ge 0
\, .
\end{equation}

\subsection*{D. Consequences}

In order to see more the linear algebra behind, we introduce now the bilinear form
$$
\langle u , v \rangle_f := \int_{I} f^2(\rho,z) \, u(\rho,z) \, v(\rho,z) \,d\rho \, dz \, ,
\qquad u,v:I \to \R \,, \quad \mbox{in} \ L^2(I)
\, ,
$$
and the induced semi-norm $\|.\|_f := \left( \langle . \, , . \rangle_f \right)^{1/2}.$ 
With this definition we can write inequality~(\ref{main}) as
$$
\langle A, \Omega\rangle_f \ge \|A\|_f^2
\, ,
$$
and immediately see that especially
\begin{equation}
\label{ieq1}
\langle \Omega, A \rangle_f = \langle A, \Omega \rangle_f \ge 0
\, .
\end{equation}
Furthermore, using the Cauchy-Schwarz inequality, we have
$\|A\|_f\|\Omega\|_f \ge \langle A, \Omega\rangle_f \ge \|A\|^2_f,$ 
and hence (since $A \equiv 0 \iff \Omega \equiv 0$) we get (in full general!) the
main result of these section, namely,
\begin{equation}
\label{ieq2}
0 \le \|A\|_f \le \|\Omega\|_f
\, ,
\end{equation}
i.e.
\begin{prop}\hfill
\\
\label{prop:gen-bounds}
\begin{equation}
\label{gen-bounds}
\mbox{\fbox{\parbox{4.8cm}{\centerline{
$\displaystyle
\rule[-2.5ex]{0mm}{6ex} 
0 \le \int_{I} f^2 \, A^2  \le  \int_{I} f^2 \; \Omega^2 $}}}}
\end{equation}
\end{prop}
(without any restriction concerning the rotation law,
$\Omega \mapsto {\cal F}(\Omega)$ in (\ref{cond:dr}), in the differentially rotating case, 
nor the regularity and sign uniformity of $\Omega \,$!). 
These integrals can be regarded as some kind of ``mean value'' with respect to the ``density'' $f^2,$ 
thus, (\ref{gen-bounds}) fulfilling the physical expectations \cite{Thorne}.

\vspace{2mm}

In addition, multiplying inequality (\ref{ieq2}) by $\|\Omega\|_f,$ 
we find (again using the Cauchy-Schwarz inequality)
$\|\Omega\|^2_f \ge \|\Omega\|_f \|A\|_f  \ge \langle\Omega, A\rangle_f,$ i.e.
\begin{equation}
\label{ieq3}
\langle \Omega, \Omega - A\rangle_f \ge 0
\, .
\end{equation}
Remarkably, the integral given in (\ref{ieq3}) has an important physical meaning;
it is, up to a constant factor, the called {\em total rotational energy} 
(see e.g.~Refs.~15,16, and 18),
$$
T \equiv \frac{1}{2} \int_{I} \Omega \; d J 
\; = \; \frac{1}{2} \int_{I} 2 \pi \,\Omega \, T_{\phi}^{\; t} \, (-{\rm g})^{\! 1/2} d\rho \, dz
\; = \; \frac{1}{16} \langle \Omega, \Omega - A\rangle_f 
$$
(also denoted $E_{\mathrm{rot}}$ or $M_{\mathrm{rot}}).$ Thus, (\ref{ieq3}) shows $T \ge 0.$
Furthermore, 
$$
16 \; T = \langle \Omega, \Omega - A \rangle_f 
= \|\Omega \|_f^2 - \langle \Omega, A \rangle_f \, \le \, \|\Omega \|_f^2,
$$
by (\ref{ieq1}). Hence, 
\begin{prop}\hfill
\\
\label{prop:T-bounds}
\begin{equation}
\label{T-bounds}
\mbox{\fbox{\parbox{9cm}{\centerline{
$\displaystyle
\rule[-2.5ex]{0mm}{6ex} 
0 \; \le \; T \equiv \frac{1}{16} \int_{I} f^2 \; \Omega \, (\Omega - A) 
\; \le \; \frac{1}{16} \int_{I} f^2 \; \Omega^2$ \, .}}}}
\end{equation}
\end{prop}
This generalizes the result given by Hartle (cf.~Ref.~7, Sec.~IV) 
in the limit of slow (differential) rotation to the general differentially rotating case.

\section*{\normalsize VI. CONCLUSIONS}

Aiming to derive general properties of equilibrium non-singular stellar models 
with differential rotation, we have established that for a wide class of rotation laws
the distribution of angular velocity of the fluid has a sign, and then both the dragging rate
(angular velocity of locally non-rotating observers) and the angular momentum density 
have the sign of the fluid angular velocity (Sec.~IV). 
In addition, the mean value (with respect to a density function) 
of the dragging rate is shown to be less than the mean value of the fluid angular velocity; 
and this is proved in full general, without having to restrict the rotation law, 
nor the uniformity in sign of the fluid angular velocity.
A further simple calculation of linear algebra
on this inequality yields a generalization of the result given by Hartle\cite{H4}
concerning positivity and upper bound of the total rotational energy
in the limit of slow (differential) rotation to the general differentially rotating case (Sec.~V).

\section*{\normalsize ACKNOWLEDGMENT}
I would like to thank F.J. Chinea and U.M. Schaudt for reading the 
manuscript and for many valuable discussions.
This work was supported by Direcci\'on General de Ense\~nanza Superior,
Project No. PB98-0772.

\appendix
\setcounter{equation}{0}
\renewcommand{\theequation}{A \arabic{equation}}

\section*{\normalsize APPENDIX A: \  Maximum (minimum) principles for 
classical sub-(super-) solutions}

By $G$ we denote an open and connected set, i.e.~a domain, in $\R^n,$ $n \ge 2.$ 
The boundary is denoted by $\partial G \equiv \ol{G} \cap (\R^n \setminus G).$
We define the differential operators
\begin{eqnarray*}
L_0 u & := & a_{ij}(x) \, \partial_{ij} u \, + \, b_i(x) \, \partial_i u \, , 
\qquad a_{ij} = a_{ji} \, ,\\
\\
\mbox{and} \quad L u & := & L_0 u \, + \, c(x) \cdot u 
\end{eqnarray*}
(where the summation convention that repeated indices
indicate summation from $1$ to $n$ is followed),
such that \cite{fn3}

\begin{enumerate}

\item[1.] $L$ (and, hence, $L_0)$ is uniformly elliptic in $G$ 
in the special form 
\begin{equation}
\label{unif_ell}
0 < \lambda \, |y|^2 \le a_{ij}(x) \, y_i y_j \le \Lambda \, |y|^2,
\quad \forall y \in \R^n \setminus \{0\}, \ \ \forall x \in G 
\qquad (|y|^2 := \sum_i y_i^{\ 2}) \, ,
\end{equation}
where $\lambda$ and $\Lambda$ are constants such that 
$0 < \lambda \le \Lambda < \infty;$ 

\item[2.] all coefficients in $L$ (and in $L_0),$ \ $a_{ij},$ $b_i$ 
(for all $i$ and $j),$ and $c,$ are measurable
and bounded functions in $G,$ 
\begin{equation}
\label{bG}
|a_{ij}| < \infty, \quad
|b_i| < \infty, \quad |c| < \infty \ \ \mbox{in} \ G \quad
(i,j \in \{1,\ldots,n\}) \, .
\end{equation}
\end{enumerate}

\begin{thm}[the weak maximum (minimum) principle for $L_0$ $(c=0) \, $]
\label{thm:wMP_L0} \hfill\\
Suppose that $L_0 u \ge 0$ $(\le 0)$ in a bounded domain $G,$
with \ $u \in C^2(G) \cap C^0(\ol{G}).$\\
Then the maximum (minimum) of $u$ is attained on the boundary,
that is,
$$
\max_{\ol{G}} u = \max_{\partial G} u \quad 
\left( \min_{\ol{G}} u = \min_{\partial G} u \right) \, .
$$
\end{thm}
(A proof of that theorem can be found e.g. in Ref.~22, Theorem~3.1.)

\bigskip 

\noindent {\bf Definition}\\
\noindent {\it For a set $G \subset \R^n,$ the boundary $\partial G$
is said to satisfy an {\rm \bf interior (exterior) sphere condition}
at a point $x_1 \in \partial G$ iff
there exists a ball $B \subset G$ \ $(B \subset \R^n \setminus \ol{G})$
\ with $x_1 \in \partial B$}

\begin{thm}[the boundary-point lemma]
\label{thm:bound-point} \hfill\\
Suppose that \ $L_0 u \ge 0$ \ $(c=0)$ in a domain $G$ not necessarily bounded.\\
Let $x_1 \in \partial G$ be such that\\
{\em (i)} \ \ \ $u$ is continuous at $x_1,$\\
{\em (ii)} \ \ $u(x_1) \ge u(x)$ for all $x \in G,$ and\\
{\em (iii)} \ $\partial G$ satisfies an {\em interior sphere condition}
at $x_1.$\\
Then the outer normal derivative of $u$ at $x_1,$ if it exists, satisfies 
the strict inequality 
$$
\partial_{{\boldsymbol \nu}} u (x_1) > 0,
$$
unless $u \equiv$ {\rm const.}$\, =u(x_1).$

\vspace{2mm}

\noindent {\em (A proof of that result can be found e.g. in Ref.~23, Theorem~7, Chap.~2.)}

\vspace{3mm}

\noindent If $c\le 0$ (in $Lu \ge 0$ ), the same conclusion holds provided \/
$u(x_1) \ge 0.$

\vspace{2mm}

\noindent {\em (See Ref.~23, Theorem~8, Chap.~2. \ Also Ref.~22, Lemma~3.4.)}
\end{thm}

\begin{thm}[the strong maximum (minimum) principle for $L$]
\label{thm:stMP_L} \hfill\\
Let $Lu \ge 0$ $(\le 0)$ in a domain $G$ not necessarily bounded, 
with \ $u \in C^2(G) \cap C^0(\ol{G}),$ and the operator $L$ satisfying
\begin{equation}
c \le 0  \quad \mbox{in} \ G
\label{c<0}
\end{equation}
apart from conditions {\em (\ref{unif_ell})} and {\em (\ref{bG})} above.\\
Then 
$u$ cannot attain a non-negative maximum (non-positive minimum) at an interior 
point of $G,$ unless $u \equiv$ {\rm const.} in $G.$

\vspace{1mm}

\noindent For $c=0,$ i.e.~$L=L_0,$ the same conclusion holds without 
the requirement `non-negative' (`non-positive').
\end{thm}
(For the proof we refer again to Ref.~22, Theorem~3.5; or Ref.~23,
Theorems~5 and 6, Chap.~2.)

\bigskip

\setcounter{equation}{0}
\renewcommand{\theequation}{B \arabic{equation}}

\section*{\normalsize APPENDIX B: \ Maximum (minimum) principle for 
generalized sub-(super-) solutions}

Consider in a domain (open and connected set) $G \subset \R^n$ \ 
$(n \ge 2)$ the differential operator with principal part of 
divergence form, defined by
$$
L u = \partial_{i} [a_{ij}(x) \partial_{j} u + a_i(x) \, u]
+ b_i(x) \, \partial_{i} u + c(x) \, u \, ,
$$
with $a_{ij}=a_{ji}.$ Notice, an operator $L$ of the general form 
$L u = \tilde{a}_{ij}(x) \partial_{ij} u + \tilde{b}_i(x) \partial_i u + \tilde{c}(x) u$ 
may be written in divergence form provided its principal coefficients $\tilde{a}_{ij}$ are
differentiable. If furthermore the $\tilde{a}_{ij}$ are constants, then 
even with coinciding coefficients $(a_{ij} = \tilde{a}_{ij},$ $b_i = \tilde{b}_i,$ 
$c = \tilde{c})$ and $a_i \equiv 0.$ \ Let us assume that
\begin{enumerate}

\item[1.] $L$ is strictly elliptic in $G,$ i.e. 
$\exists$ a constant $\lambda > 0$ such that $\lambda \le$
the minimum eigenvalue of the principal coefficient matrix $[a_{ij}(x)],$ 
\begin{equation}
\label{st_ell}
\lambda \, |y|^2 \le a_{ij}(x) \, y_i y_j \ \ \ \ \forall y \in \R^n,
\ \ \forall x \in G \, ;
\end{equation}
\item[2.] $a_{ij},$ $a_i,$ $b_i,$ and $c$ are measurable
and bounded functions in $G,$ 
\begin{equation}
\label{b}
|a_{ij}| < \infty, \quad |a_i| < \infty, \quad
|b_i| < \infty, \quad |c| < \infty \ \ \mbox{in} \ G \quad
(i,j \in \{1,\ldots,n\}) \, .
\end{equation}
\end{enumerate}

By definition, for a function $u$ which is only assumed to be {\em weakly differentiable}   
and such that the functions \, $a_{ij} \partial_{j} u + a_i u$ \, and \, 
$b_i \partial_{i} u + c u,$ $i=1,\ldots, n$ \ are locally integrable (in particular, for 
$u$ belonging to the Sobolev space $W^{1,2}(G)),$ \ $u$ is said to satisfy
$Lu = g$ in $G$ {\em in a generalized (or weak) sense} 
$(g$ also a locally integrable function in $G)$
if it satisfies
\begin{eqnarray*}
{\cal L} (u, \varphi; G) & := & \int_{G} \{ (a_{ij} \partial_{j} u + a_i u) 
\partial_{i} \varphi - (b_i \partial_{i} u + c u) \varphi \}
d x \\
& = & - \int_{G} g \, \varphi \, d x, \ \ \ \ \
\forall \varphi \ge 0 \ \ \ \varphi \in C^1_c(G) \, 
\end{eqnarray*}
(where $C^1_c(G)$ is the set of functions in $C^1(G)$ with compact support in $G).$

\vspace{1mm}

Notice, $u$ is {\em generalized sub-(super-)solution} relative to a
differential operator $L$ and the domain $G$ (i.e.~satisfies $Lu \ge 0$ $(\le 0)$ in $G$ 
in a generalized sense) if it satisfies 
${\cal L} (u, \varphi; G) \le 0$ $(\ge 0), \ \ \
\forall \varphi \ge 0 \ \ \varphi \in C^1_c(G).$

\begin{thm}[strong maximum (minimum) principle]
\label{thm:stMPg} \hfill\\
Let $u \in W^{1,2}(G) \cap C^0(G)$ satisfy $Lu \ge 0$ $(\le 0)$ in $G$ in a generalized sense,
with 
\begin{equation}
\int_{G} (c \varphi - a_i \partial_i \varphi) \, d x \le 0, \ \ \  
\forall \varphi \ge 0 \ \ \ \varphi \in C^1_c(G) \, .
\label{c<0_g}
\end{equation}
(equivalent to requirement {\em (\ref{c<0})} in the classical case)
and conditions {\em (\ref{st_ell})} and {\em (\ref{b})} above.\\
Then
$u$ cannot achieve a non-negative maximum (non-positive minimum) in the interior of $G,$ 
unless $u \equiv$ {\rm const.}
\end{thm}
(A proof of this theorem can be found in Ref.~22, Theorem~8.19.)




\begin{thebibliography}{99}
\bibitem{fn3} 
There are more general forms of maximum principles, i.e.~with much 
more relaxed conditions (see e.g. Refs.~22 and 23), 
however these are not needed for the present application.
\end{thebibliography}
\end{document}